\documentclass[longauth]{aa}  

\usepackage{graphicx}
\usepackage{txfonts}
\usepackage{gensymb}
\usepackage[colorlinks=true,linkcolor=blue,citecolor=blue,urlcolor=blue]{hyperref}
\newcommand{\prot}{$P_{\text{rot}}$}
\newcommand{\mps}{m s$^{-1}$}
\newcommand{\gcm}{g cm$^{-3}$}
\newcommand{\teff}{$T_{\text{eff}}$}
\defcitealias{kostov19}{K19}

\begin{document}

   \title{Characterization of the L 98-59 multi-planetary system with HARPS}
      \subtitle{Two confirmed terrestrial planets and a mass upper limit on the third
     \thanks{Based on observations made with the HARPS instrument 
       on the ESO 3.6 m telescope under the program 
       IDs 198.C-0838(A), 1102.C-0339(A), and 0102.C-0525 at Cerro La Silla (Chile).}
     %\textsuperscript{,}\thanks{Data (Tables XXX) are only 
     %  available at the CDS via anonymous ftp to 
     %  cdsarc.u-strasbg.fr (130.79.128.5)\newline
     %  or via\newline
     %  http://cdsarc.u-strasbg.fr/viz-bin/qcat?J/A+A/XXX/XXX
     %}
   }

   \authorrunning{Cloutier et al.}
   \titlerunning{HARPS characterization of the L 98-59 planetary system}

   \author{R. Cloutier\inst{\ref{toronto1},\ref{toronto2},\ref{montreal}} \and
    N. Astudillo-Defru\inst{\ref{conce1}} \and
     X. Bonfils\inst{\ref{grenoble}} \and
     J. S. Jenkins\inst{\ref{uchile}} \and
     G. Ricker\inst{\ref{mit1}} \and
     R. Vanderspek\inst{\ref{mit1}} \and
     D. W. Latham\inst{\ref{cfa}} \and
     S. Seager\inst{\ref{mit2}} \and
     J. Winn\inst{\ref{princeton}} \and
     J. M. Jenkins\inst{\ref{nasa}} \and
     J. M. Almenara\inst{\ref{grenoble}} \and
     F. Bouchy\inst{\ref{geneva}} \and
     X. Delfosse\inst{\ref{grenoble}} \and
     M. R. D\'iaz\inst{\ref{uchile}} \and
     R. F. D\'iaz\inst{\ref{uba},\ref{iafe}} \and
     R. Doyon\inst{\ref{montreal}} \and
     P. Figueira\inst{\ref{eso},\ref{caup}} \and
     T. Forveille\inst{\ref{grenoble}} \and
     T. Jaffe\inst{\ref{nasagsfc}} \and
     N. T. Kurtovic\inst{\ref{uchile}} \and
     C. Lovis\inst{\ref{geneva}} \and
     M. Mayor\inst{\ref{geneva}} \and
     K. Menou\inst{\ref{toronto1},\ref{toronto2}} \and
     E. Morgan\inst{\ref{mit1}} \and
     R. Morris\inst{\ref{seti},\ref{nasa}} \and
     P. Muirhead\inst{\ref{boston}} \and
     F. Murgas\inst{\ref{grenoble}} \and
     F. Pepe\inst{\ref{geneva}} \and
     N. C. Santos\inst{\ref{caup}, \ref{porto}} \and
     D. S\'egransan\inst{\ref{geneva}} \and
     J. C. Smith\inst{\ref{seti},\ref{nasa}} \and
     P. Tenenbaum\inst{\ref{seti},\ref{nasa}} \and
     G. Torres\inst{\ref{cfa}} \and
     S. Udry\inst{\ref{geneva}} 
     M. Vezie\inst{\ref{mit1}} \and
     J. Villasenor\inst{\ref{mit1}}
   }

   \institute{Dept. of Astronomy \& Astrophysics, University of Toronto, 50 St. George Street, M5S 3H4, Toronto, ON, Canada \label{toronto1}
     \and Centre for Planetary Sciences, Dept. of Physical \& Environmental Sciences, University of Toronto Scarborough, 1265 Military Trail, M1C 1A4, Toronto, ON, Canada \label{toronto2}
    \and Institut de Recherche sur les Exoplan\`etes, D\'epartement de Physique, Universit\'e de Montr\'eal, Montr\'eal QC, H3C 3J7, Canada \label{montreal}
    \and Departamento de Matem\'atica y F\'isica Aplicadas, Universidad Cat\'olica de la Sant\'isima Concepci\'on, Alonso de Rivera 2850, Concepci\'on, Chile \label{conce1}
    \and Univ. Grenoble Alpes, CNRS, IPAG, F-38000 Grenoble, France \label{grenoble}
    \and Departamento de Astronom\'ia, Universidad de Chile, Camino El Observatorio 1515, Las Condes, Santiago, Chile \label{uchile}
    \and Kavli Institute for Astrophysics and Space Research, Massachusetts Institute of Technology, Cambridge, MA 02139, USA \label{mit1}
    % \and Department of Terrestrial Magnetism, Carnegie Institution for Science, 5241 Broad Branch Road NW, Washington DC 20015, USA\label{carnegiedtm}
     %\and Observatories of the Carnegie Institution for Science, 813 Santa Barbara Street, Pasadena, CA 91101 \label{carnegieobs}
     %\and Instituto de Astrof\'isica, Facultad de F\'isica, Pontificia Universidad Cat\'olica de Chile, Av. Vicuña Mackenna 4860, 7820436 Macul, Santiago, Chile \label{puc}
     %\and Millennium Institute of Astrophysics, 7820436 Santiago, Chile \label{mas}
     %\and Astrophysics Group, Keele University, Staffordshire, ST5 5 BG, UK\label{Keele}
     \and Harvard-Smithsonian Center for Astrophysics, 60 Garden Street, Cambridge, MA 02138, USA \label{cfa}
     \and Department of Earth, Atmospheric, and Planetary Sciences, Massachusetts Institute of Technology,  Cambridge,  MA 02139, USA \label{mit2}
     \and Department of Astrophysical Sciences, Princeton University, Princeton, NJ 08544, USA \label{princeton}
     \and NASA  Ames  Research  Center,  Moffett  Field,  CA  94035, USA \label{nasa}
     \and Observatoire de Gen\`eve, Universit\'e de Gen\`eve, 51 ch. des Maillettes, 1290 Sauverny, Switzerland \label{geneva}
     \and Universidad de Buenos Aires, Facultad de Ciencias Exactas y Naturales. Buenos Aires, Argentina \label{uba}
     \and CONICET - Universidad de Buenos Aires. Instituto de Astronom\'ia y F\'isica del Espacio (IAFE). Buenos Aires, Argentina \label{iafe}
     \and European Southern Observatory, Alonso de C\'ordova 3107, Vitacura, Regi\'on Metropolitana, Chile \label{eso}
     \and Instituto de Astrof\'isica e Ci\^encias do Espa\c{c}o, Universidade do Porto, CAUP, Rua das Estrelas, PT4150-762 Porto, Portugal \label{caup}
     \and NASA Goddard Space Flight Center, Greenbelt, MD USA \label{nasagsfc}
    \and SETI Institute, Mountain View, CA 94043, USA \label{seti}
     \and Department of Astronomy \& Institute for Astrophysical Research, Boston University, 725 Commonwealth Avenue, Boston, MA 02215, USA \label{boston}
    \and Departamento de F\'isica e Astronomia, Faculdade de Ci\^encias, Universidade do Porto, Portugal \label{porto}
    %\and American Association of Variable Star Observers, 49 Bay State Road, Cambridge, MA 02138, USA \label{aavso}
    %\and Physics Department and Tsinghua Centre for Astrophysics, Tsinghua University, Beijing 100084, China \label{tsinghua}
    %\and Computational Engineering and Science Research Centre, University of Southern Queensland, Toowoomba, QLD, 4350, Australia \label{cesrc}
    %\and Max-Planck-Institut fur Astronomie, K\"unigstuhl 17, D-69117 Heidelberg, Germany \label{mpia}
    %\and Campo Catino Astronomical Observatory, Regione Lazio, Guarcino (FR), 03010 Italy \label{ccao}
    %\and Department of Physics and Astronomy, University of Louisville, Louisville, KY 40292, USA \label{louisville}
    %\and Departamento de F\'isica, Universidade Federal do Rio Grande do Norte, 59078-970 Natal, RN, Brazil \label{riogrande}
    %\and Departamento de Astronom\'ia, Universidad de Concepci\'on, Casilla 160-C, Concepci\'on, Chile \label{conce2}
    %\and Leidos, Inc., Moffett Field, CA 94035, USA \label{leidos}
    %\and Department of Physics and Astronomy, Vanderbilt University, Nashville, TN 37235, USA \label{Vanderbilt} 
    %\and Center for Space Research, MIT, 37-414, Cambridge, MA 02139, UA \label{mit3}
    }

   \date{}

 \abstract{}{L 98-59 (TIC 307210830, TOI-175) is a nearby M3 dwarf around which TESS revealed three terrestrial-sized transiting planets (0.80, 1.35, 1.57 Earth radii) in a compact configuration with orbital periods shorter than 7.5 days. Here we aim to measure the masses of the known transiting planets in this system using precise radial velocity (RV) measurements taken with the HARPS spectrograph.}{We consider both trained and untrained Gaussian process regression models of stellar activity to simultaneously model the RV data with the planetary signals. Our RV analysis is then supplemented with dynamical simulations to provide strong constraints on the planets' orbital eccentricities by requiring long-term stability.}{We measure the planet masses of the two outermost planets to be $2.46\pm 0.31$ and $2.26\pm 0.50$ Earth masses which confirms their bulk terrestrial compositions. We are able to place an upper limit on the mass of the smallest, innermost planet of $<0.98$ Earth masses with 95\% confidence. Our RV + dynamical stability analysis places strong constraints on the orbital eccentricities and reveals that each planet's orbit likely has $e<0.1$ to ensure a dynamically stable system.}{The L 98-59 compact system of three likely rocky planets offers a unique laboratory for studies of planet formation, dynamical stability, and comparative atmospheric planetology. Continued RV monitoring will help refine the characterization of the innermost planet and potentially reveal additional planets in the system at wider separations.} 
  
   %\abstract{hello}


   \maketitle
%-------------------------------------------------------------------
\section{Introduction}
NASA's \emph{Transiting Exoplanet Survey Satellite} \citep[TESS;][]{ricker15} is expected to discover thousands of new transiting planetary systems around nearby stars over $\sim 80$\% of the entire sky \citep{sullivan15,ballard19,barclay18,huang18b}. Throughout its 2-year long primary mission, TESS will observe $\gtrsim 200,000$ targets from the TESS Input Catalog \citep[TIC;][]{stassun17} at a 2 minute cadence as well as many more targets within the 30 minute full frame images. Indeed several confirmed planetary systems have already been uncovered by TESS within its first year of operations \citep{brahm18,jones18,canas19,dragomir19,espinoza19,kipping19,kostov19,neilsen19,quinn19,rodriguez19,wang19,vanderspek19} including a small number of planets that have begun to contribute to the completion of the mission's level one science requirement of delivering the masses of 50 planets smaller than 4 R$_{\oplus}$ ($\pi$ Mensae c; \citealt{gandolfi18,huang18a}, TOI-402.01, 02; \citealt{dumusque19}).

The nearby M3 dwarf L 98-59 (TIC 307210830, TOI-175, d=10.6 pc, Table~\ref{tab:star}) was included in the TESS Input Catalog based on its stellar parameters from the Cool Dwarf list \citep{muirhead18} and so far has been observed in TESS Sector 2. Three terrestrial-sized planetary candidates around L 98-59 (TOI-175.01: $P_1$=3.69 days, $r_{p,1}=1.35$ R$_{\oplus}$, TOI-175.02: $P_2$=7.45 days, $r_{p,2}=1.57$ R$_{\oplus}$, TOI-175.03: $P_3$=2.25 days, $r_{p,3}=0.80$ R$_{\oplus}$) were flagged by the Science Processing Operations Center Pipeline \citep[SPOC;][]{jenkins16} and subsequently passed a set of validation tests \citep{twicken18,li19} prior to being published as TESS Data Alerts\footnote{\url{https://tess.mit.edu/alerts/}}. Many of the properties of this multi-planet system make it of interest for radial velocity (RV) mass characterization \citep{cloutier18b}, planetary atmospheric characterization \citep{kempton18,louie18}, and direct investigations of M dwarf planet formation, evolution, and system architectures \citep{lissauer11,fabrycky14}. As such, the system warranted an intensive follow-up campaign presented by  \citealt{kostov19} (hereafter \citetalias{kostov19}) that ruled out astrophysical false positive scenarios and confirmed the planetary nature of each of the three planet candidates. 

In this paper we present the results of our follow-up study to obtain precise planet masses for as many of the L 98-59 planets as possible using HARPS precision RVs. In practice we are only able to recover robust masses for the two outermost planets TOI-175.01 and 02 but we also report our derived upper limits on the mass of the smallest known planet TOI-175.03. In Sect.~\ref{sec:rv} we discuss our spectroscopic HARPS observations, in Sect.~\ref{sec:model} we establish our model of the observed RVs before presenting our results in Sect.~\ref{sec:results}. In Sect.~\ref{sec:stability} we use the measured planet masses to perform a dynamical stability analysis of the system to provide stronger constraints on each planet's orbital eccentricity. We then conclude with a discussion in Sect.~\ref{sec:conclusion}.

\begin{table}[t]
  \caption{L 98-59 stellar parameters.}
  \label{tab:star}
  \centering
  \small
  \begin{tabular}{lcc}  
    \hline\noalign{\smallskip}
    Parameter & Value & Reference \\
    \hline\noalign{\smallskip}
    \multicolumn{3}{c}{\emph{L 98-59, TIC 307210830, TOI-175}} \\
    \noalign{\smallskip}
    \multicolumn{3}{c}{\emph{Astrometry}} \\
    RA, $\alpha$ [deg] & $124.532860$ & 1,2 \\
    Dec, $\delta$ [deg] & $-68.314466$ & 1,2 \\
    RA proper motion, $\mu_{\alpha}$ && \\
    $[$mas yr$^{-1}]$ & $94.767\pm 0.054$ & 1,2 \\
    Dec proper motion, $\mu_{\delta}$ && \\
    $[$mas yr$^{-1}]$ & $-340.470\pm 0.052$ & 1,2 \\
    Parallax, $\varpi$ [mas] & $94.167\pm 0.028$ & 1,2,3 \\
    Distance, $d$ [pc] & $10.619\pm 0.003$ & 1,2,3 \\
    \noalign{\smallskip}
    \multicolumn{3}{c}{\emph{Photometry}} \\
    $B$ & $13.289\pm 0.027$ & 4 \\
    $g'$ & $12.453\pm 0.019$ & 4 \\
    $V$ & $11.685\pm 0.017$ & 4 \\
    $r'$ & $11.065\pm 0.044$ & 4 \\
    $G_{BP}$ & $11.977\pm 0.002$ & 1,5 \\
    $G$ & $10.5976\pm 0.0008$ & 1,5 \\
    $G_{RP}$ & $9.472\pm 0.001$ & 1,5 \\
    $T$ & $9.393$ & 6 \\
    $J$ & $7.933\pm 0.027$ & 7 \\
    $H$ & $7.359\pm 0.049$ & 7 \\
    $K_s$ & $7.101\pm 0.018$ & 7 \\
    $W_1$ & $6.935\pm 0.062$ & 8 \\
    $W_2$ & $6.767\pm 0.021$ & 8 \\
    $W_3$ & $6.703\pm 0.016$ & 8 \\
    $W_4$ & $6.578\pm 0.047$ & 8 \\
    \noalign{\smallskip}
    \multicolumn{3}{c}{\emph{Stellar parameters}} \\
    Spectral type & M3V $\pm 1$ & 9 \\
    Stellar radius, $R_s$ [R$_{\odot}$]$^*$ & $0.314\pm 0.014$ & 3,9 \\
    Effective temperature, \teff{} [K] & $3412\pm 49$ & 3,10 \\
    Stellar mass, $M_s$ [M$_{\odot}$]$^{\dagger}$ & $0.312\pm 0.031$ & 3,11 \\
    Surface gravity, $\log{g}$ [dex] & $4.94\pm 0.06$ & 12 \\
    Metallicity, [Fe/H] & $-0.5\pm 0.5$ & 9 \\
    $\log{R'_{HK}}$ & $-5.40\pm 0.11$ & 12 \\
    Rotation period, \prot{} [days]$^{\ddagger}$ & $78\pm 13$ & 12,13 \\
    %Projected rotation velocity & & \\
    %$v\sin{i}$ [km s$^{-1}$] & $<1.9$ & 9 \\
    \hline\noalign{\smallskip}
  \end{tabular}
  
  \begin{list}{}{}
    \item 1) \citealt{gaia18}, 2) \citealt{lindegren18}, 3) \citealt{cloutier19b}, 4) \citealt{henden16}, 5) \citealt{evans18}, 6) \citealt{stassun17}, 7) \citealt{cutri03}, 8) \citealt{cutri13}, 9) \citealt{kostov19}, 10) \citealt{mann15}, 11) \citealt{benedict16}, 12) this work, 13) \citealt{astudillodefru17b}.
    \item $^*$ Includes the radius uncertainty from \citep{kostov19}. 
    \item $^{\dagger}$ We add in quadrature a fractional uncertainty of 10\% based on the dispersion in stellar masses for stars with metallicities that differ from solar \citep{mann19}.
    % see their section 7.5
    \item $^{\ddagger}$ The predicted \prot{} based on $\log{R'_{HK}}$ and the M dwarf activity-rotation relation from \cite{astudillodefru17b}. This value is consistent with the periodogram peak in the $H\alpha$ time series in Fig.~\ref{fig:GLSP}.
  \end{list}
\end{table}

%--------------------------------------------------------------------
\section{HARPS Observations}
\label{sec:rv}
\subsection{HARPS data acquisition}
Using the \textit{High Accuracy Radial velocity Planet Searcher} \citep[HARPS;][]{mayor03} echelle spectrograph mounted at the 3.6m ESO telescope at La Silla Observatory, Chile, we obtain a set of 161 spectra of L 98-59 between October 17, 2018 (BJD = 2458408.5) and April 28, 2019 (BJD = 2458601.5). The HARPS optical spectrograph at $R=115,000$ is stabilized in pressure and temperature which helps enable its sub-\mps{} accuracy. 

Throughout the 5-month observing campaign of L 98-59 we elected not to use a simultaneous wavelength calibration 
(ie. on-sky calibration fibre) to prevent possible contamination of the bluer spectral orders by the calibration lamp. In the ESO programs 198.C-0838 and 1102.C-0339 (140/161 observations) the exposure time was set to 900 seconds, resulting in a median signal-to-noise ratio (S/N) of 41 per resolution element at 650 nm and a median measurement uncertainty of 1.61 \mps{.} In the ESO program 0102.C-0525 (21/161 observations) the exposure time ranges between 500 seconds and 1800 seconds, with a median S/N of 49 per resolution element at 650 nm and a median measurement uncertainty of 2.08 \mps{.} 

\subsection{Radial velocity extraction}
To compute the RV time series we performed a maximum likelihood analysis between a stellar
template and individual spectra following \citet{astudillodefru17a}. The adopted stellar template corresponds
to the median of all spectra that were previously shifted to the star frame. A telluric template was 
derived by the median of spectra that were shifted to the Earth frame. For these two steps we used the 
stellar radial velocity derived by the HARPS Data Reduction Software \citep[DRS;][]{lovis07} through a cross-correlation
function. We used the barycentric Earth radial velocity as computed by the DRS as well. The resulting stellar template 
was Doppler shifted over a window of 40 km/s wide and centered on the average of the RVs 
computed by the DRS (-5.661 km/s). The telluric template was used to mask the spectral zones contaminated 
by telluric lines. For each RV step we computed the value of the likelihood function with the maximum of the 
likelihood
function representing the RV of the spectrum under analysis. The process was repeated for the entire HARPS
dataset and resulted in the RV time series reported in Table~\ref{tab:timeseries} that is used in the subsequent
analysis.

\begin{table*}[t]
\caption{HARPS spectroscopic time series.}
\label{tab:timeseries}
\centering
\small
\begin{tabular}{ccccccccccccccc}
\hline\noalign{\smallskip}
Time & RV$^{*}$ & $\sigma_{\text{RV}}$ & $H\alpha$ & $\sigma_{H\alpha}$ & $H\beta$ & $\sigma_{H\beta}$ & $H\gamma$ & $\sigma_{H\gamma}$ & NaD & $\sigma_{NaD}$ & S-index & $\sigma_{S}$ & FWHM & BIS \\ 
$[$BJD - & [\mps{]} & [\mps{]} & $\times 10^2$ & $\times 10^2$ & $\times 10^2$ & $\times 10^2$ & $\times 10^2$ & $\times 10^2$ & $\times 10^2$ & $\times 10^2$ & - & - & $[$km s$^{-1}]$ & $[$km s$^{-1}]$ \\ 
2,457,000$]$ &&&&&&&&&&&&&& \\ 
\hline\noalign{\smallskip}
1408.853661 & -5678.7 & 2.3 & 6.93 & 0.02 & 5.43 & 0.05 & 12.32 & 0.17 & 0.91 & 0.02 & 0.69 & 0.09 & 3.0588 & 23.2798 \\ 
1409.844622 & -5678.9 & 2.4 & 6.77 & 0.02 & 5.22 & 0.05 & 11.46 & 0.17 & 0.84 & 0.02 & 0.65 & 0.10 & 3.0637 & 23.3301 \\ 
1412.858886 & -5679.1 & 2.2 & 7.07 & 0.02 & 5.79 & 0.05 & 12.85 & 0.17 & 0.98 & 0.02 & 0.80 & 0.10 & 3.0559 & 23.3108 \\ 
1413.860798 & -5676.2 & 1.9 & 6.79 & 0.02 & 5.25 & 0.04 & 11.87 & 0.13 & 0.83 & 0.02 & 0.65 & 0.07 & 3.0655 & 23.3935 \\ 
1414.858984 & -5674.6 & 2.5 & 6.81 & 0.02 & 5.12 & 0.06 & 11.75 & 0.19 & 0.85 & 0.03 & 0.71 & 0.12 & 3.0560 & 23.1147 \\ 
\hline\noalign{\smallskip}
\end{tabular}

\begin{list}{}{}
\item \textbf{Notes.} Only the first five rows are depicted here for clarity.
\item $^{*}$ Systemic velocity, $\gamma_0 = -5678.4\pm 0.2$ \mps{.}
\end{list}

\end{table*}

\section{Model Setup}
\label{sec:model}

\subsection{Periodogram analysis}
\label{subsec:kep}
To identify strong periodicities in our RV time series we compute the generalized Lomb-Scargle periodogram \citep[GLSP;][]{zechmeister09} of all spectroscopic time series derived from our HARPS spectra and of its window function (WF). The ancillary spectroscopic time series of $H\alpha$, $H\beta$, $H\gamma$, the sodium doublet NaD, and the S-index based on the Ca H \& K doublet are sensitive to chromospheric activity and may therefore be used to identify periodicities in the RV data arising from chromospheric activity sources such as plages.

The $H\alpha$ index was computed using the same pass-bands as \cite{gomesdasilva12}, that is a band of 1.6 \AA\ wide centered on 6562.8 \AA\ and two control bands of widths 10.75 \AA\ and 8.75 \AA\
centered on 6550.87 \AA\ and 6580.31 \AA, respectively; the central 
band for $H\beta$ was limited by 4861.04 \AA--4861.60 \AA, and we defined two control bands limited by
4855.04 \AA--4860.04 \AA\ and 4862.6 \AA--4867.2 \AA; for $H\gamma$ we integrated over three bands
bounded by 4333.60 \AA--4336.80 \AA, 4340.16 \AA--4340.76 \AA, and 4342.00 \AA--4344.00 \AA; 
the central bands to calculate the NaD-index were similar as 
\citet{gomesdasilva12}, namely wide of 0.5 \AA\ and centered on 5889.95 \AA\ and 5895.92 \AA, 
but with control bands limited by 5860.0 \AA--5870.0 \AA\ and 5904.0 \AA--5908.0 \AA; 
the S-index was calculated following \cite{duncan1991} and the calibration derived in
\citet[][Eq.~3]{astudillodefru17b} that scales the index computed from HARPS spectra to Mount Wilson.

Similarly, the full width at half maximum (FWHM) and bisector (BIS) shape parameters of the spectral CCF may be
sensitive to chromospheric and/or photospheric active regions such as dark spots \citep{queloz01,desort07}.
Periodicities in either the FWHM or BIS time series may therefore also allude to periodic signals arising from
stellar activity. We analyzed the FWHM and BIS as derived by the HARPS DRS. The GLSP of the WF is also computed
to potentially identify sources of aliasing from our time sampling.

\begin{figure}
    \centering
    \includegraphics[width=0.98\hsize]{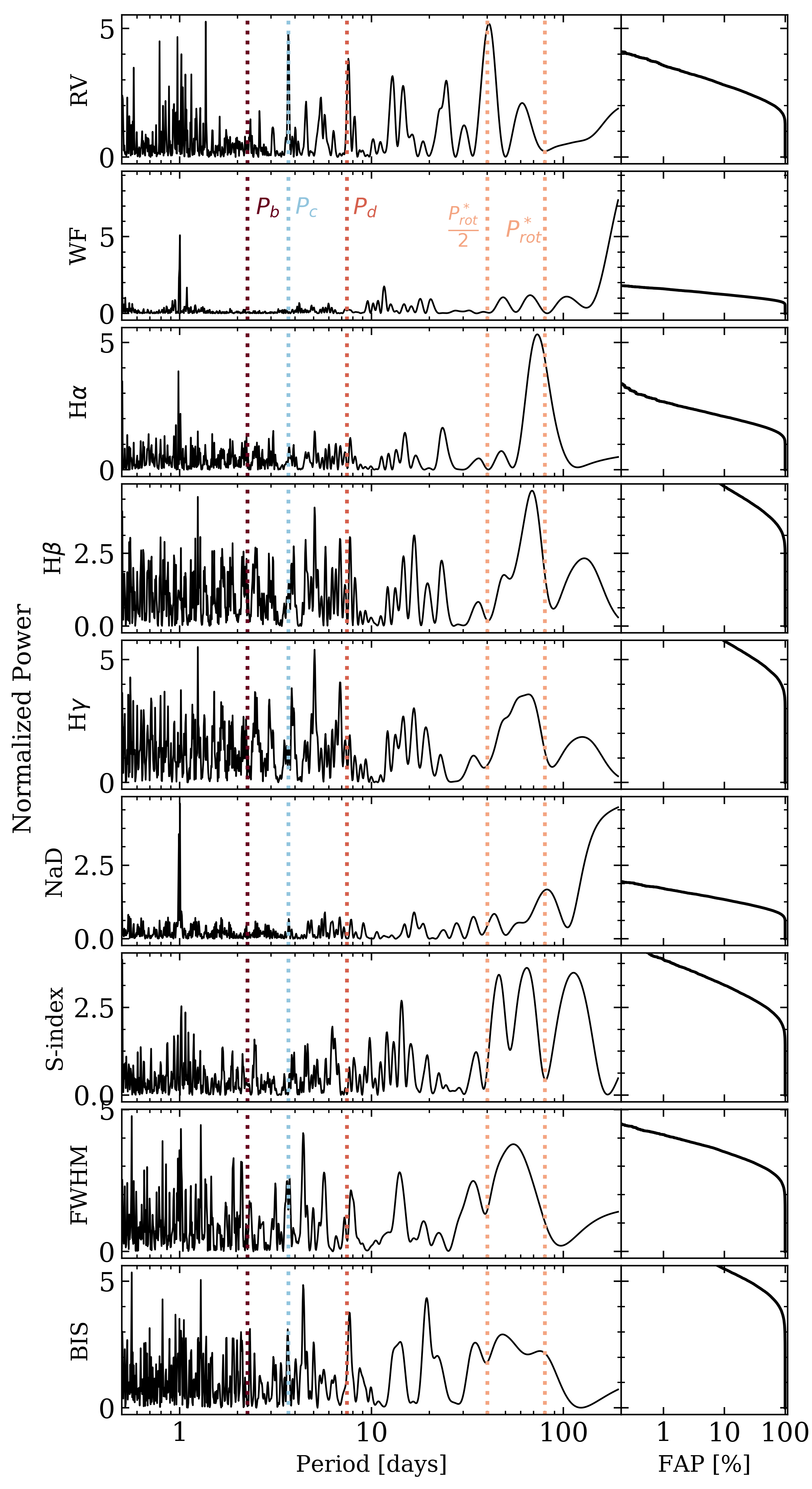}
    \caption{\emph{Left column}: generalized Lomb-Scargle periodograms of our HARPS RV time series, its window function, and the $H\alpha$, $H\beta$, $H\gamma$, sodium doublet, S-index, full width half maximum, and bisector activity indicators. The vertical dotted lines highlight the orbital periods of the three known transiting planets plus the posited rotation period of L 98-59 at $P_{\text{rot}} \sim 80$ days and its first harmonic $P_{\text{rot}}/2$. \emph{Right column}: the false alarm probabilities computed from bootstrapping with replacement.}
    \label{fig:GLSP}
\end{figure}

The resulting GLSPs are shown in Fig.~\ref{fig:GLSP} along with their false alarm probability (FAP) curves. The FAP curves are computed via bootstrapping with replacement using $10^4$ iterations and normalizing each GLSP's power scale by its standard deviation. The GLSP of the RVs reveals a moderately significant peak close to the orbital period of L 98-59d ($\sim 7.45$ days) plus significant peaks at the orbital period of L 98-59c ($\sim 3.69$ days) as well as a broad peak centered around $\sim 40$ days. The photometric rotation period of L 98-59 remains undetected in the SAP TESS light curve and furthermore the star exhibits a negligible rotational broadening \citepalias[$v\sin{i} < 1.9$ km s$^{-1}$;][]{kostov19} indicative of L 98-59 being a largely inactive, old M dwarf with a likely long rotation period and a correspondingly low amplitude of photometric variability \citep{newton16a}. However, the strongest periodic signal as seen in any activity sensitive time series in Fig.~\ref{fig:GLSP} is a broad feature centered around $\sim 80$ days in the $H\alpha$ GLSP (FAP $\ll 0.1$\%). This signal is consistent with the expected \prot{} $=78\pm 13$ days based on the star's value of $\log{R'_{HK}}=-5.4\pm 0.11$ and using the M dwarf magnetic activity-rotation relation from \cite{astudillodefru17b}.
If this signal is indeed due to stellar rotation at \prot{} $\sim 80$ days then it could explain the $\sim 40$ day signal in the RV GLSP as being the first harmonic of \prot{.} Nearly all of the remaining activity sensitive time series have GLSPs that are consistent with noise (i.e. FAP $\gtrsim 10$\%) implying that no strong periodic signals are resolved in those time series. Two notable exceptions exist. The first is the broad peak at $\gtrsim 100$ days in the NaD time series that is also seen in the GLSP of the WF. The second is a less significant peak that is intermediate between 40 and 80 days and persists in each of the $H\beta$, $H\gamma$, S-index, and FWHM GLSPs. The origin of this weak, intermediate peak is unknown but may also be related to stellar rotation as the $\sim 80$ day $H\alpha$ peak is posited to be.

A periodic signal from the remaining planet L 98-59b ($\sim 2.25$ days) is not seen at a low FAP in the GLSP of the raw RVs (Fig.~\ref{fig:GLSP}). However our iterative periodogram analysis will reveal the presence of this signal, as well as the increased strength of the L 98-59d planetary signal ($\sim 7.45$ days), upon the joint modelling of the planets with RV stellar activity in Sect.~\ref{sec:results}.

\subsection{Stellar activity}
\label{subsec:act}
Stellar activity on M dwarfs predominantly arises from active regions in the stellar photosphere and chromosphere \citep{lindegren03}. The resulting RV signal is modulated by stellar rotation as active regions traverse the visible stellar disk and disrupt its symmetry thus creating a temporally correlated RV variation that can mask or even  mimic planetary signals under certain circumstances \citep{vanderburg16}. In this study, we will use the $H\alpha$ activity indicator, which is unaffected by planet-induced Doppler shifts, to inform our stellar activity model. 

Following its successful application to activity modelling in M dwarf planetary systems \citep[e.g.][]{astudillodefru17c,cloutier17b,bonfils18,cloutier19,ment19}, we adopt a semi-parametric Gaussian process (GP) regression model of RV stellar activity to simultaneously model activity with the RV planetary signals. Given a parameterization of the temporal covariance structure in our time series, the semi-parametric nature of the GP activity treatment is well-suited to modelling a stochastic physical process like stellar activity without requiring a deterministic functional form. Here we assume that the apparent stellar activity signal seen in the RV data at $\sim 40$ days has a manifestation in the $H\alpha$ data at $\sim 80$ days and whose temporal covariance structure is quasi-periodic as it is rotationally modulated and yet it is not purely periodic due to evolution in the active region sizes, contrasts, and spatial distribution over multiple rotation cycles \citep{giles17}. The corresponding covariance kernel function of our GP model, trained on the $H\alpha$ time series, is 

\begin{equation}
    k(t_i,t_j) = a^2 \exp{\left[ -\frac{(t_i-t_j)^2}{2\lambda^2} -\Gamma^2 \sin^2{\left( \frac{\pi |t_i-t_j|}{P_{\text{rot}}} \right)} \right]}
    \label{eq:kernel}
\end{equation}

\noindent and is parameterized by the following GP hyperparameters: the covariance amplitude $a$, an exponential decay time scale $\lambda$, a coherence parameter $\Gamma$, and the strong periodic signal seen in the $H\alpha$ GLSP (Fig.~\ref{fig:GLSP}) which we attribute to the L 98-59 rotation period \prot{.} 

We sample the posterior probability density function (PDF) of the logarithmic GP hyperparameters by running the $H\alpha$ time series through a Markov chain Monte Carlo (MCMC) simulation using the \texttt{emcee} ensemble sampler package \citep{foremanmackey13}. The prior PDFs on each of the GP hyperparameters are reported in Table~\ref{tab:priors}. The ln likelihood function used to sample their joint posterior PDF is given by 

\begin{equation}
    \ln{\mathcal{L}} = -\frac{1}{2} \left( \mathbf{y}^{\text{T}}\cdot \mathbf{K}\cdot \mathbf{y} + \ln{\text{det}\mathbf{K}} + N\ln{2\pi} \right)
\end{equation}

\noindent where $\textbf{y}$ is the vector of $N$ $H\alpha$ measurements taken at times $\mathbf{t}=\{t_1,t_2,\dots,t_N \}$ and the $N \times N$ covariance matrix $\mathbf{K}$ is given by

\begin{equation}
    \mathbf{K}_{ij} = k(t_i,t_j) + \delta_{ij}(\boldsymbol{\sigma}_{H\alpha}(t_i)^2 + s^2).
\end{equation}

\noindent The inclusion of the Kronecker delta $\delta_{ij}$ term adds the $H\alpha$ measurement uncertainties $\boldsymbol{\sigma}_{H\alpha}$ to the diagonal elements of $\mathbf{K}$ and includes an additive jitter factor $s$. Hence the full set parameters sampled during the $H\alpha$ training phase is $\{ \ln{a}, \ln{\lambda}, \ln{\Gamma}, \ln{P_{\text{rot}}}, s \}$.

\begin{table}[t]
  \caption{L 98-59 model parameter priors.}
  \label{tab:priors}
  \centering
  \small
  \begin{tabular}{lc}  
    \hline\noalign{\smallskip}
    Parameter & Prior \\
    \hline\noalign{\smallskip}
    \multicolumn{2}{c}{$H\alpha$ training model} \\
    \noalign{\smallskip}
    ln covariance amplitude, $\ln{a}$ & $\mathcal{U}(-10,-2)$ \\
    ln exponential time scale, $\ln{\lambda}$ & $\mathcal{U}(1,20)$ \\
    ln coherence, $\ln{\Gamma}$ & $\mathcal{U}(-10,10)$ \\
    ln rotation period, $\ln{P_{\text{rot}}}$ & $\mathcal{U}(2,5)$ \\
    Additive jitter, $s$ & $\mathcal{J}(10^{-6},10^{-3})$ \\
    \hline\noalign{\smallskip}
    \multicolumn{2}{c}{RV model} \\
    \noalign{\smallskip}
    ln covariance amplitude, $\ln{a}$ & $\mathcal{U}(-5,5)$ \\
    ln exponential time scale, $\ln{\lambda}$ & $p(\ln{\lambda}|H\alpha)$ \\
    ln coherence, $\ln{\Gamma}$ & $p(\ln{\Gamma}|H\alpha)$ \\
    ln periodic time scale, $\ln{P_{\text{GP}}}$ & $p(\ln{[P_{\text{rot}}/2]}|H\alpha)$ \\
    Additive jitter, $s$ [\mps{]} & $\mathcal{J}(10^{-2},10)$ \\
    Systemic velocity, $\gamma_0$ [\mps{]} & $\mathcal{U}(-10,10)$ \\
    \noalign{\smallskip}
    \multicolumn{2}{c}{\emph{L 98-59b (TOI-175.03)}} \\
    Orbital period, $P_b$ [days] & $\mathcal{N}(2.2532, 3\times 10^{-4})$ \\
    Time of mid-transit, $T_{0,b}$ & \\
    $[$BJD - 2,457,000$]$ & $\mathcal{N}(1366.1708, 1\times 10^{-4})$ \\
    Semi-amplitude, $K_b$ [\mps{]} & $\mathcal{J}(0.1,10)$ \\
    $h_b=\sqrt{e_b}\cos{\omega_b}^{*}$ & $\mathcal{U}(-1,1)$ \\
    $k_b=\sqrt{e_b}\sin{\omega_b}^{*}$ & $\mathcal{U}(-1,1)$ \\    \noalign{\smallskip}
    \multicolumn{2}{c}{\emph{L 98-59c (TOI-175.01)}} \\
    Orbital period, $P_c$ [days] & $\mathcal{N}(3.6904, 2.5\times 10^{-4})$ \\
    Time of mid-transit, $T_{0,c}$ & \\
    $[$BJD - 2,457,000$]$ & $\mathcal{N}(1367.2751, 6\times 10^{-4})$ \\
    Semi-amplitude, $K_c$ [\mps{]} & $\mathcal{J}(0.1,10)$ \\
    $h_c=\sqrt{e_c}\cos{\omega_c}^{*}$ & $\mathcal{U}(-1,1)$ \\
    $k_c=\sqrt{e_c}\sin{\omega_c}^{*}$ & $\mathcal{U}(-1,1)$ \\
    \noalign{\smallskip}
    \multicolumn{2}{c}{\emph{L 98-59d (TOI-175.02)}} \\
    Orbital period, $P_d$ [days] & $\mathcal{N}(7.4513, 7\times 10^{-4})$ \\
    Time of mid-transit, $T_{0,d}$ & \\
    $[$BJD - 2,457,000$]$ & $\mathcal{N}(1362.7375, 8\times 10^{-4})$ \\
    Semi-amplitude, $K_d$ [\mps{]} & $\mathcal{J}(0.1,10)$ \\
    $h_d=\sqrt{e_d}\cos{\omega_d}^{*}$ & $\mathcal{U}(-1,1)$ \\
    $k_d=\sqrt{e_d}\sin{\omega_d}^{*}$ & $\mathcal{U}(-1,1)$ \\
    
    \hline\noalign{\smallskip}
  \end{tabular}
  
  \begin{list}{}{}
  \item $^{*}$ We also require that the corresponding eccentricity value $e_i<1$.
  \end{list}
\end{table}

Because we assume that the physical process that gives rise to variations in the $H\alpha$ time series has a manifestation in the observed RVs, we can use the constraints on the GP hyperparameters from training to inform our RV model of stellar activity. In particular, the posterior PDFs of the covariance parameters $\{ \ln{\lambda}, \ln{\Gamma}, \ln{P_{\text{rot}}} \}$ will be used throughout our modelling of the stellar RVs to 
derive self-consistent activity and planetary solutions with minimal contamination of the latter by the former as a result of training.

\subsection{Radial velocity model}
\label{sec:rvmodel}
Following the training of the GP activity model on the $H\alpha$ time series we can proceed with modelling the RVs. Our RV model contains four physical components from stellar activity plus the three known transiting planets around L 98-59. The RV GP activity model features the same covariance function as was adopted during training (Eq.~\ref{eq:kernel}) and therefore contains the five hyperparameters $\{ a, \lambda, \Gamma, P_{\text{GP}}, s \}$ where $a$ and $s$ are unique to the RVs whereas the priors on the remaining hyperparameters $\lambda$, $\Gamma$ and $P_{\text{GP}}$ are constrained by their joint posterior PDF from training. Recall that the apparent rotation signal in $H\alpha$ at \prot{} $\sim 80$ days appears to be manifested at its first harmonic (\prot{}$/2$) in the RVs at $\sim 40$ days (see Fig.~\ref{fig:GLSP}). As such, we modify the marginalized posterior on \prot{} from training by rescaling $P_{\text{GP}} \to P_{\text{rot}}/2$ and use the modified PDF as a prior on $P_{\text{GP}}$ in our RV model.

The three planetary signals in our RV model are treated as independent keplerian orbital solutions. This simplification neglects any gravitational interactions between the planets and makes the sampling of the planetary parameter posterior PDFs much more computationally tractable by negating the need to run dynamical simulations at every step in the MCMC chains.
We can justify this simplification by first noting that the planets do not appear to exhibit significant transit timing variations (TTVs) at the level of precision for which such TTVs would be resolvable with the TESS photometric precision \citepalias[$\sim 5.1$ minutes for L 98-59b;][]{kostov19}. Furthermore, \citetalias{kostov19} performed long-term dynamical simulations of the system by considering the maximum a-posteriori (MAP) planet mass predictions and their $+1\sigma$ values for each of the known L 98-59 planets. The planet mass predictions were based on the planets' measured radii and the use of the  \texttt{forecaster} tool \citep{chen17}\footnote{The MAP predicted planet masses for L 98-59b, c and d are 0.3, 2.0, and 2.3 M$_{\oplus}$ respectively. Their MAP $+1\sigma$ predicted masses are 0.5, 3.6, and 4.2 M$_{\oplus}$.}. Combining each planet's predicted mass with their osculating orbital elements, and assuming initially circular orbits, the orbital eccentricities of the three planets remained nearly circular (i.e. $\lesssim 0.006$) after one million orbits of the outermost planet (i.e. $\sim 21$ thousand years). In contrast, \citetalias{kostov19} reported that half of the simulations with initial eccentricities of 0.1 became unstable. These results support the notion that the orbits of the L 98-59 planets are nearly circular, a result that itself is consistent with other compact multi-planet systems exhibiting low eccentricities \citep[$\lesssim 0.05$;][]{hadden14,vaneylen15}.

As a back of the envelope calculation, the difference between the RV semi-amplitudes of the three L 98-59 planets with circular orbits and with $e=0.1$ (assuming the MAP predicted planet masses) is $<1$ cm s$^{-1}$ for all planets. Even for eccentricities of $e=0.23$, for which orbit crossing would occur for both planet pairs, the difference in RV semi-amplitudes between that and the circular orbit scenario is $<5$ cm s$^{-1}$. These discrepancies between the circular and maximally elliptical system architectures are well below the typical HARPS RV measurement uncertainty of 2.06 \mps{} such that the differences in the planet-induced stellar RV curves from the superposition of keplerian solutions and from N-body integrations, for which non-zero eccentricities would develop, are negligible. By adopting the simplification of keplerian orbits, in our MCMC we are effectively sampling the orbital parameters of each planet's osculating orbit rather than tracking the time evolution of those orbital parameters due to mutual planetary interactions. This simplification holds given that the dynamical variations in those orbits are small compared to the level of precision of our data.

%we conduct an N-body dynamical simulation of the L 98-59 system and compute the resulting stellar RVs sampled using our HARPS WF. Using the same input parameters we also compute the stellar RVs from the superposition of three keplerian orbits; i.e. non-interacting planets. The N-body and keplerian RV curves are directly compared in Fig.~\ref{fig:RVcurvecomp} and reveals that the maximum difference between the two curves is \textbf{X} \mps{.} The rms of the difference between the two curves is \textbf{X} \mps{.} Both of these values are well below the median RV measurement uncertainty of our HARPS time series of $\sigma_{\text{RV}}=\mathbf{X}$ \mps{} and therefore justifies our use of the more computationally tractable keplerian superposition when modelling the observed L 98-59 RV variations.

Our complete RV model therefore includes the five GP hyperparameters of stellar activity, the L 98-59 systemic velocity $\gamma_0$ plus five keplerian parameters for each known planet. Namely, each planet's orbital period $P_i$ ($i$ in the planet index; $i=b,c,d$), time of mid-transit $T_{0,i}$, RV semi-amplitude $K_i$, $h_i=\sqrt{e_i}\cos{\omega_i}$, and $k_i=\sqrt{e_i}\sin{\omega_i}$ \citep{ford06} where $e_i$ and $\omega_i$ are the planet's orbital eccentricity and argument of periastron respectively. Our complete RV model therefore contains 21 parameters $\{\ln{a}, \ln{\lambda}, \ln{\Gamma}, \ln{P_{\text{GP}}}, s, \gamma_0, P_b, T_{0,b}, K_b, h_b, k_b, P_c, T_{0,c}, K_c, h_c, k_c,$ $P_d, T_{0,d},$ $K_d, h_d, k_d \}$. We adopt Gaussian priors on each planet's orbital period and time of mid-transit based on the results of their transit light curve analysis \citepalias{kostov19} as those data have much more constraining power on the planet ephemerides than do the RVs alone. We adopt broad uninformative priors on the remaining keplerian parameters which are reported in Table~\ref{tab:priors}.

\section{Results}
\label{sec:results} 
\begin{figure*}
    \centering
    \includegraphics[width=.99\hsize]{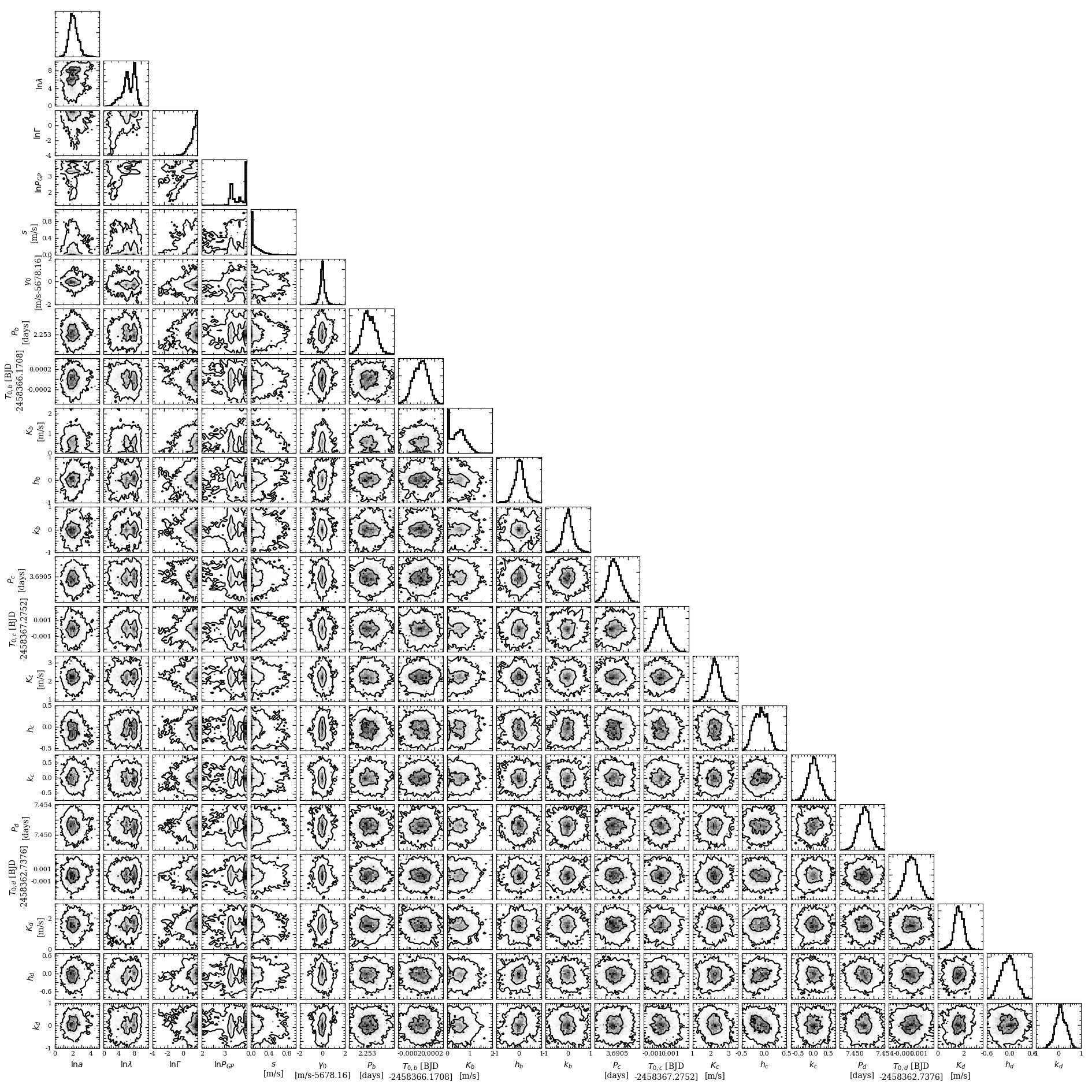}
    \caption{The marginalized and joint posterior probability density functions of the model parameters from our L 98-59 RV analysis. The adopted RV model includes a trained GP activity model ($\{ \ln{a}, \ln{\lambda}, \ln{\Gamma}, \ln{P_{\text{GP}}}, s \}$) plus the star's systemic velocity ($\{ \gamma_0 \}$) and three keplerian planet solutions ($\{ P_i, T_{0,i}, K_i, h_i, k_i \}$ for $i=b,c,d$).}
    \label{fig:corner}
\end{figure*}

The resulting joint and marginalized posterior PDFs from our MCMC analysis are shown in Fig.~\ref{fig:corner}. Unless stated otherwise, point estimates of each parameter correspond to their MAP values and are reported in Table~\ref{tab:results} along with their uncertainties from the 16th and 84th percentiles of their marginalized posterior PDF.

The marginalized posterior PDF of the $\ln{P_{\text{GP}}}$ hyperparameter does not have a well-defined solution that was expected from the training phase. Furthermore, the $\ln{\Gamma}$ posterior PDF is highly asymmetric and clearly favours larger values than were favoured by the $H\alpha$ time series. Although we are not principally interested in the values of the RV GP hyperparameters, their values might exhibit a direct effect on the planetary parameters. To investigate this we also consider an alternative model consisting of the three planets plus an \emph{untrained} GP activity model. This analysis is carried out identically to when using the trained GP except that the priors on $\{ \ln{\lambda}, \ln{\Gamma}, \ln{P_{\text{GP}}} \}$ are modified to the following uninformative priors: $\mathcal{U}(1,20)$, $\mathcal{U}(-10,10)$, and $\mathcal{U}(0,5)$ respectively. The resulting point estimates of the model parameters are also reported in Table~\ref{tab:results} and we find that all keplerian planet parameters are consistent at the $1\sigma$ level between the two models considered. 

We also estimated the Bayesian evidence $\mathcal{Z}$ for each RV model featuring a trained and untrained GP activity component respectively. The evidences were computed using the \cite{perrakis14} estimator and the marginalized posterior PDFs from our MCMC analyses as importance samplers. This evidence estimator has been shown to result in quantitatively similar results to other more robust but computationally expensive methods (e.g. nested samplers; \citealt{nelson18}). The resulting Bayes factor, or evidence ratio, between competing models containing a trained and untrained GP activity component is 0.3 thus indicating that inferences resulting from either model are nearly equivalent. Following the consistency of the two models considered we opt to focus on the results from the trained RV model in the subsequent analysis and discussion.

\begin{figure*}
    \centering
    \includegraphics[width=.9\hsize]{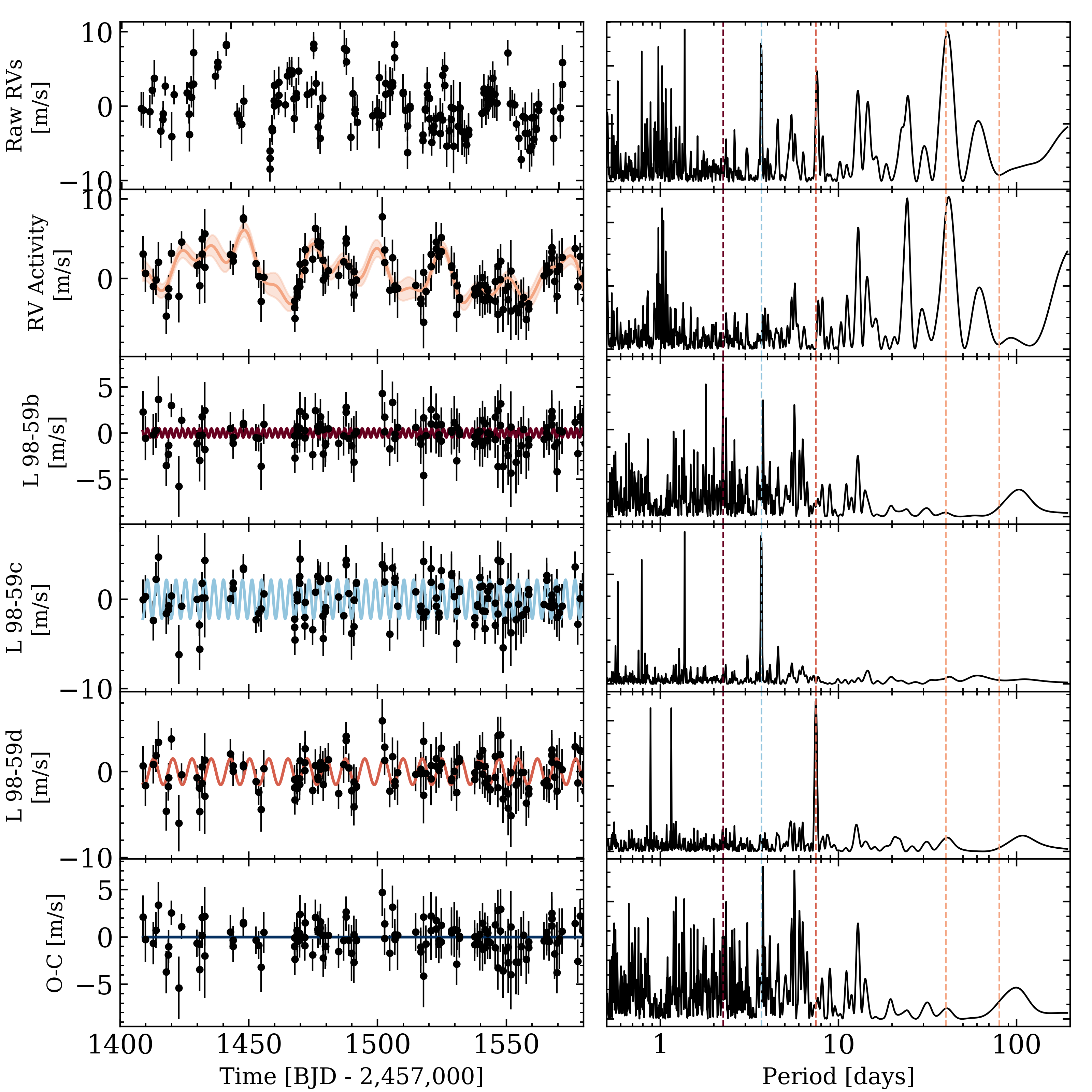}
    \caption{RV time series (\emph{left column}) of each physical component included in our model of the L 98-59 RVs along with its GLSP (\emph{right column}). In addition to the raw RVs (\emph{top panel}), also depicted is the RV activity (\emph{second panel}), the three L 98-59 planets (\emph{third, fourth, and fifth panels}), and the RV residuals (\emph{bottom panel}). The RV activity model depicted is the mean GP function along with its $\pm 1\sigma$ uncertainty in the surrounding shaded region. The L 98-59c and d planet curves are their MAP keplerian orbital solutions while the L 98-59b curve is its median keplerian orbit.}
    \label{fig:rvcomponents}
\end{figure*}

Fig.~\ref{fig:rvcomponents} depicts each of the MAP components of our RV model along with its corresponding GLSP. The stellar activity component (i.e. the RVs less the three keplerian solutions) has a maximum amplitude of $\sim 6$ m/s and is dominated by the clear periodicity at $\sim 40$ days that was seen in the GLSP of the raw RVs. Removal of the mean GP activity model mitigates that signal at $\sim 40$ days. The planetary RV components from L 98-59c and d are each dominated by their known orbital periods with some aliasing at shorter orbital periods that are consequently mitigated once the planet's MAP keplerian solution is subtracted off. The keplerian model of the remaining planet L 98-59b has a small median RV semi-amplitude of $K_b = 0.48$ \mps{} making its orbital period only slightly resolved in its GLSP. Lastly, the GLSP of the RV residuals is consistent with noise indicating that we have modelled all major sources of RV variation in our HARPS time series.

\begin{figure}
    \centering
    \includegraphics[width=.96\hsize]{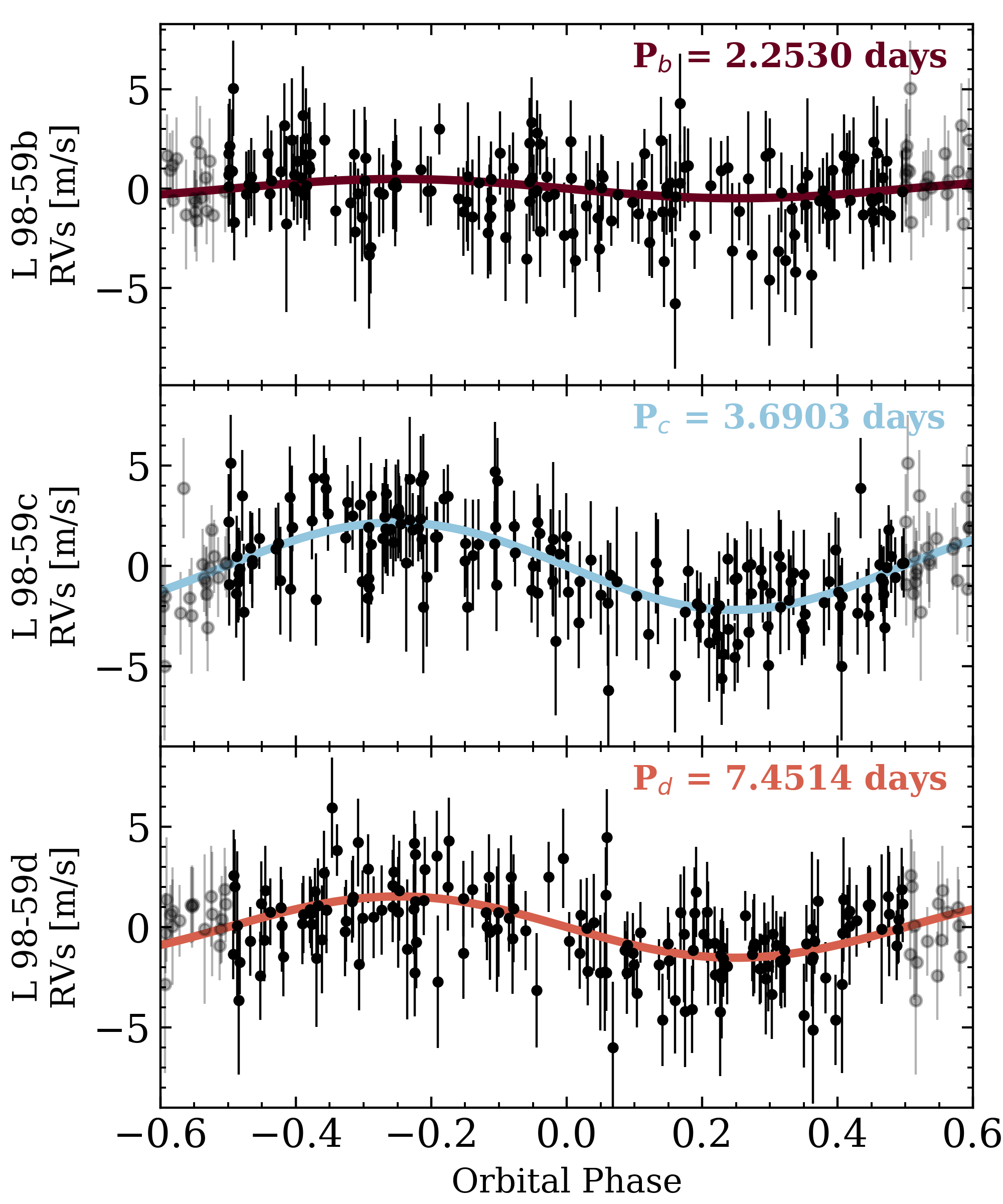}
    \caption{Phase-folded RVs for each known planets in the L 98-59 system. Each set of RVs has been corrected for stellar activity and the two planets not depicted in its panel. Only the two outermost planets are detected with semi-amplitudes that are inconsistent with 0 \mps{.}}
    \label{fig:rvphased}
\end{figure}

Only the two outer planets have RV semi-amplitude `detections' in that their measured values are robustly $>0$ \mps{.} These values are $K_c=2.21\pm 0.28$ \mps{} and $K_d=1.61\pm 0.36$ \mps{} and represent 7.9 and $4.5\sigma$ detections respectively. The marginalized posterior PDF of $K_b$, corresponding to the smallest and innermost planet in our model, has a median value of 0.48 \mps{} but is consistent with 0 \mps{} therefore resulting in a non-detection of $K_b$ in our dataset. Instead we are only able to place an upper limit on $K_b$ of $<1.03$ at 95\% confidence. The phase-folded RVs are shown in Fig.~\ref{fig:rvphased} along with the MAP L 98-59c and d keplerian orbital solutions and the median L 98-59b keplerian orbit. The periodic modulation from L 98-59c and d are clearly discernible. Meanwhile the median value of $K_b$ corresponds to a $\lesssim 1.5\sigma$ detection and is not discernible in the phase-folded RVs.

\begin{figure}
    \centering
    \includegraphics[width=\hsize]{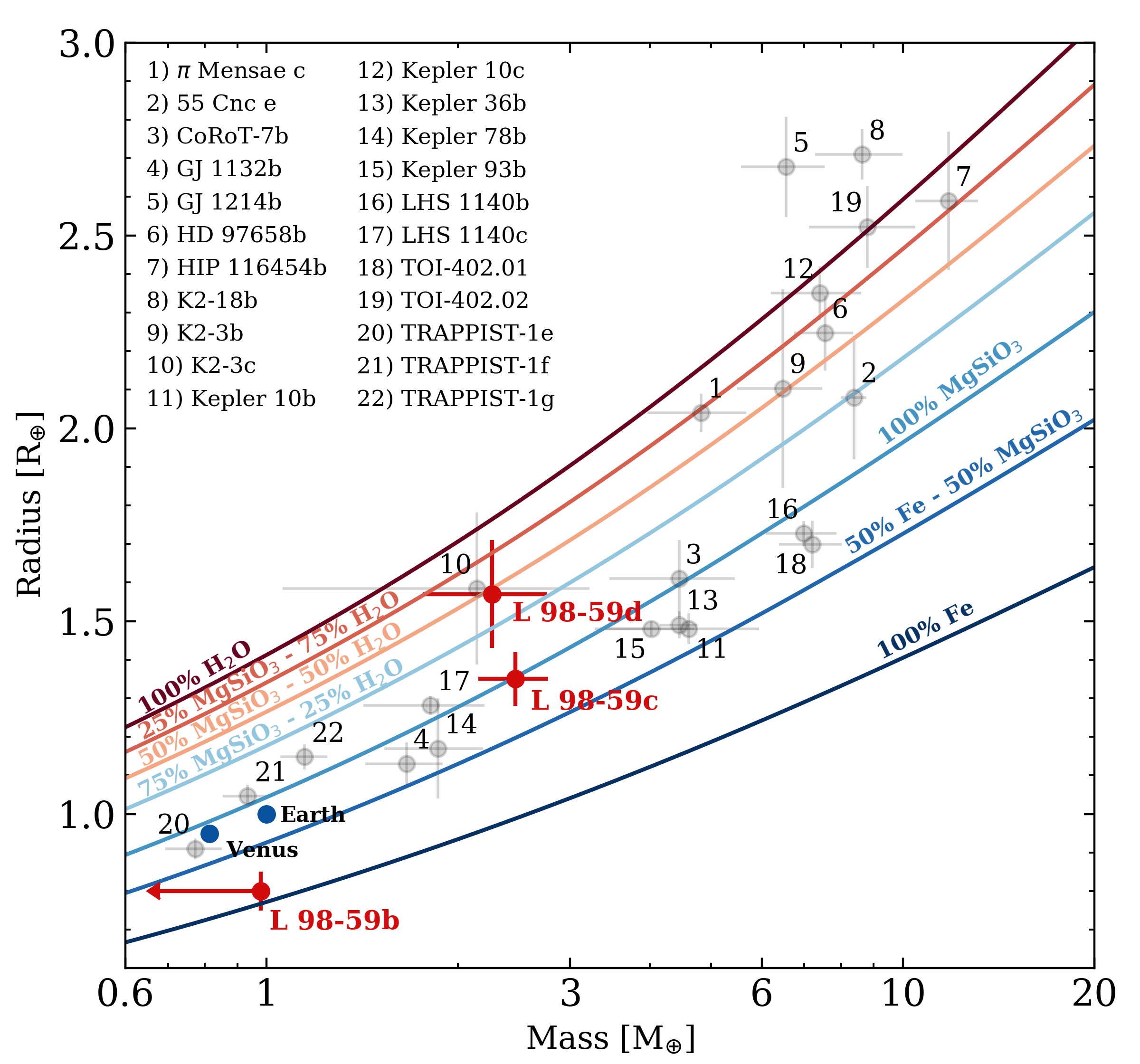}
    \caption{The three L 98-59 planets in the planetary mass and radius space along with a set of small exoplanets, Earth, and Venus for comparison. The solid lines depict theoretical mass-radius curves from two component models of fully differentiated planetary interiors with fractional compositions by mass in water (H$_2$O), rock (MgSiO$_3$), and/or iron (Fe) \citep{zeng13}. Each model's interior composition is annotated above its respective curve.}
    \label{fig:MR}
\end{figure}

The planetary masses corresponding to the RV semi-amplitudes measured with our data are $m_{p,b} < 0.98$ M$_{\oplus}$ (at 95\% confidence), $m_{p,c}=2.46\pm 0.31$ M$_{\oplus}$, and $m_{p,d}=2.26\pm 0.50$ M$_{\oplus}$. \citetalias{kostov19} measure planet-star radius ratios ($r_{p,b}/R_s=0.0234\pm 0.0009$, $r_{p,c}/R_s=0.0396\pm 0.0010$, and $r_{p,d}/R_s=0.0462\pm 0.0029$) from which they derive planetary radii of $r_{p,b}=0.80\pm 0.05$ R$_{\oplus}$, $r_{p,c}=1.35\pm 0.07$ R$_{\oplus}$, and $r_{p,d}=1.57\pm 0.14$ R$_{\oplus}$. The measured planetary masses and radii result in constraints on the planets' bulk densities of $\rho_{p,b}<12.7$ \gcm{,} $\rho_{p,c}=5.5\pm 1.2$ \gcm{,} $\rho_{p,d}=3.3\pm 1.2$ \gcm{} thus making the two outer planets consistent with bulk compositions that are dominated by rock as demonstrated in the planetary mass-radius plane in Fig.~\ref{fig:MR}. Although the non-detection of $K_b$ prevents a precise value of its bulk density from being derived, the close proximity of L 98-59b to its host star and its intermediate size between that of Earth and Mars are evidence for its terrestrial nature \citep{owen13,jin14,lopez14,chen16,lopez16,owen17}. Further constraints on $\rho_{p,b}$ may be realized as we note that the upper limit on $\rho_{p,b}<12.7$ \gcm{} from the 95\% confidence interval of the $\rho_{p,b}$ marginalized posterior exceeds the bulk density of a pure iron ball the size of L 98-59b (12.2 \gcm{;} \citealt{zeng13}). This implies that the true RV semi-amplitude of L 98-59b is likely $\lesssim 1$ \mps{} and that the detection of $K_b$ will require much more stringent RV follow-up with an instrument whose performance on L 98-59 is similar to or better than HARPS, such as ESPRESSO \citep{pepe10}.

\begin{table*}[t]
  \caption{Model parameters for the L 98-59 three planet system.}
  \label{tab:results}
  \centering
  \small
  \begin{tabular}{lcc}  
    \hline\noalign{\smallskip}
    & 3-planet model plus GP activity & 3-planet model plus an untrained \\
    & model trained on H$\alpha$ & GP activity model \\
    \hline\noalign{\smallskip}
    Systemic velocity, $\gamma_0$ [\mps{]} & $-5678.4\pm 0.2$ & $-5678.3\pm 0.3$ \\
    \noalign{\smallskip}
    
    \emph{GP hyperparameters} && \\
    \noalign{\smallskip}
    Covariance amplitude, $a$ [\mps{]} & $7.6^{+4.8}_{-2.6}$ & $6.9^{+4.0}_{-2.3}$ \\
    \noalign{\smallskip}
    Exponential timescale, $\lambda$ [days] & $770^{+3308}_{-632}$ & $1452^{+2819}_{-1359}$ \\
    \noalign{\smallskip}
    Coherence, $\Gamma$ & $4.5^{+1.9}_{-2.3}$ & $7.5^{+10.1}_{-4.9}$ \\
    \noalign{\smallskip}
    Periodic timescale, $P_{\text{GP}}$ [days] & $51.4^{+1.3}_{-24.6}$ & $51.2^{+6.5}_{-24.3}$ \\
    \noalign{\smallskip}
    Additive jitter, $s$ [\mps{]} & $0.06^{+0.18}_{-0.05}$ & $0.08^{+0.19}_{-0.07}$ \\
    \noalign{\smallskip}
    \hline
    \noalign{\smallskip}
    
    \emph{Measured parameters} && \\
    \noalign{\smallskip}
    \multicolumn{3}{c}{\emph{L 98-59b (TOI-175.03)}} \\
    Orbital period, $P_b$ [days] & \multicolumn{2}{c}{$2.2531\pm 0.0004$} \\
    Time of mid-transit, $T_{0,b}$ [BJD-2,457,000] & \multicolumn{2}{c}{$1366.1708\pm 0.0001$} \\
    Semi-amplitude, $K_b$ [\mps{]}$^*$ & $<1.03$ & $<1.07$ \\
    $h_b=\sqrt{e_b}\cos{\omega_b}$ & $0.03\pm 0.22$ &  $-0.07\pm 0.24$ \\
    $k_b=\sqrt{e_b}\sin{\omega_b}$ & $-0.01\pm 0.22$ & $-0.06\pm 0.26$ \\
    \noalign{\medskip}
    \multicolumn{3}{c}{\emph{L 98-59c (TOI-175.01)}} \\
    Orbital period, $P_c$ [days] & \multicolumn{2}{c}{$3.6904\pm 0.0003$} \\
    Time of mid-transit, $T_{0,c}$ [BJD-2,457,000] & \multicolumn{2}{c}{$1367.2752\pm 0.0006$} \\
    Semi-amplitude, $K_c$ [\mps{]} & $2.21\pm 0.28$ & $2.13\pm 0.32$ \\
    $h_c=\sqrt{e_c}\cos{\omega_c}$ & $0.08\pm 0.16$ &  $-0.05\pm 0.16$ \\
    $k_c=\sqrt{e_c}\sin{\omega_c}$ & $0.01\pm 0.19$ & $0.04\pm 0.22$ \\
    \noalign{\medskip}
    \multicolumn{3}{c}{\emph{L 98-59d (TOI-175.02)}} \\
    Orbital period, $P_d$ [days] & \multicolumn{2}{c}{$7.4512\pm 0.0007$} \\
    Time of mid-transit, $T_{0,d}$ [BJD-2,457,000] & \multicolumn{2}{c}{$1362.7376\pm 0.0009$} \\
    Semi-amplitude, $K_d$ [\mps{]} & $1.61\pm 0.36$ & $1.59\pm 0.38$ \\
    $h_d=\sqrt{e_d}\cos{\omega_d}$ & $-0.04\pm 0.20$ & $0.02\pm 0.18$ \\
    $k_d=\sqrt{e_d}\sin{\omega_d}$ & $0.08\pm 0.25$ &  $-0.01\pm 0.23$ \\
    \noalign{\smallskip}
    ln Bayesian evidence, $\ln{\mathcal{Z}}$ & -411.5 & -410.3 \\
    \noalign{\smallskip}
    \hline
    \noalign{\smallskip}
    
    \emph{Derived parameters} && \\
    \noalign{\smallskip}
    \multicolumn{3}{c}{\emph{L 98-59b (TOI-175.03)}} \\
    Semi-major axis, $a_b$ [AU] & \multicolumn{2}{c}{$0.02282 \pm 0.00008$} \\
    Equilibrium temperature, $T_{\text{eq},b}$ [K] && \\
    \hspace{10pt} Bond albedo = 0 & \multicolumn{2}{c}{$610\pm 13$} \\
    \hspace{10pt} Bond albedo = 0.3 & \multicolumn{2}{c}{$558\pm 12$} \\
    Planet radius, $r_{p,b}$ [R$_{\oplus}$]$^{\dagger}$ & \multicolumn{2}{c}{$0.80\pm 0.05$} \\
    Planet mass, $m_{p,b}$ [M$_{\oplus}$]$^*$ & $<0.98$ & $<1.01$ \\
    Bulk density, $\rho_b$ [g cm$^{-3}$]$^*$ & $<12.7$ & $<13.1$ \\
    Surface gravity, $g_b$ [m s$^{-2}$]$^*$ & $<16.6$ & $<17.1$ \\
    Escape velocity, $v_{\text{esc},b}$ [km s$^{-1}$]$^*$ & $<14.3$ & $<14.7$ \\
    Eccentricity, $e_b^{\ddagger}$ & $<0.12$ & $<0.15$ \\
    \noalign{\medskip}
    \multicolumn{3}{c}{\emph{L 98-59c (TOI-175.01)}} \\
    Semi-major axis, $a_c$ [AU] & \multicolumn{2}{c}{$0.0317\pm 0.0001$} \\
    Equilibrium temperature, $T_{\text{eq},c}$ [K] && \\
    \hspace{10pt} Bond albedo = 0 & \multicolumn{2}{c}{$517\pm 11$} \\
    \hspace{10pt} Bond albedo = 0.3 & \multicolumn{2}{c}{$473\pm 10$} \\
    Planet radius, $r_{p,c}$ [R$_{\oplus}$]$^{\dagger}$ & \multicolumn{2}{c}{$1.35\pm 0.07$} \\
    Planet mass, $m_{p,c}$ [M$_{\oplus}$] & $2.46\pm 0.31$ & $2.36\pm 0.36$ \\
    Bulk density, $\rho_c$ [g cm$^{-3}$] & $5.5\pm 1.2$ & $5.3\pm 1.2$ \\
    Surface gravity, $g_c$ [m s$^{-2}$] & $13.3\pm 2.2$ & $12.8\pm 2.4$ \\
    Escape velocity, $v_{\text{esc},c}$ [km s$^{-1}$] & $15.1\pm 1.1$ & $14.8\pm 1.2$ \\
    Eccentricity, $e_c^{\ddagger}$ & $<0.07$ & $<0.07$ \\
    \noalign{\medskip}
    \multicolumn{3}{c}{\emph{L 98-59d (TOI-175.02)}} \\
    Semi-major axis, $a_d$ [AU] & \multicolumn{2}{c}{$0.0506\pm 0.0002$} \\
    Equilibrium temperature, $T_{\text{eq},d}$ [K] && \\
    \hspace{10pt} Bond albedo = 0 & \multicolumn{2}{c}{$409\pm 8$} \\
    \hspace{10pt} Bond albedo = 0.3 & \multicolumn{2}{c}{$374\pm 8$} \\
    Planet radius, $r_{p,d}$ [R$_{\oplus}$]$^{\dagger}$ & \multicolumn{2}{c}{$1.57\pm 0.14$} \\
    Planet mass, $m_{p,d}$ [M$_{\oplus}$] & $2.26\pm 0.50$ & $2.24\pm 0.53$ \\
    Bulk density, $\rho_d$ [g cm$^{-3}$] & $3.3\pm 1.2$ & $3.2\pm 1.2$ \\
    Surface gravity, $g_d$ [m s$^{-2}$] & $9.0\pm 2.7$ & $8.9\pm 2.8$ \\
    Escape velocity, $v_{\text{esc},d}$ [km s$^{-1}$] & $13.4\pm 1.6$ & $13.4\pm 1.7$ \\
    Eccentricity, $e_d^{\ddagger}$ & $<0.09$ & $<0.08$ \\
    \noalign{\smallskip}\hline
  \end{tabular}
  
  \begin{list}{}{}
      \item $^{*}$ Upper limit given by the 95\% confidence interval.
      \item $^{\dagger}$ Planetary radii from \citetalias{kostov19}.
      \item $^{\ddagger}$ Upper limit given by the 95\% confidence interval derived from the joint RV and dynamical stability analyses.
  \end{list}
\end{table*}

\section{Dynamical Stability and Eccentricity Constraints}
\label{sec:stability}
The presence of three planets in a compact configuration around L 98-59 provides a unique opportunity to provide additional constraints on the planets' orbital eccentricities using stability criteria to limit the range of permissible eccentricities. \citetalias{kostov19} showed through dynamical simulations (assuming MAP planet mass predictions from \citealt{chen17}) that for initially circular orbits the system can be long-lived but as the initial eccentricities were increased to just 0.1, many of their simulated planetary systems became unstable in $\lesssim 20,000$ years. Using the planetary mass measurements and upper limits derived in this paper we can use dynamical simulations to constrain each planet's eccentricity given that the system must remain stable for at least the duration of the simulation.

We proceed with deriving the fraction of stable systems as a function of each planet's orbital eccentricity by simulating $10^4$ realizations of the L 98-59 planetary system and integrating each system forward in time using the \texttt{WHFast} symplectic integrator \citep{rein15} within the open-source \texttt{REBOUND} N-body package \citep{rein12}. In each realization, the stellar mass is draw from $\mathcal{N}(0.312,0.003)$ M$_{\odot}$ which in turn prescribes each planet's initial semi-major axis when combined with its orbital period that are drawn from their marginalized posterior PDF from Sect.~\ref{sec:results}. \citetalias{kostov19} noted that despite having a period ratio of 2.02, the two outer planets are likely just wide of a resonant configuration such that we do not attempt to force the outer planet pair to converge towards a mean motion resonance in our dynamical simulations. Similarly to the orbital periods, each planet's mass and orbital phase at $t=0$ are drawn from their marginalized posterior PDFs from Sect.~\ref{sec:results}. Orbital inclinations are drawn from the approximately Gaussian posterior PDFs reported in Table 2 of \citetalias{kostov19}. The argument of periastron and longitude of the ascending node for each planet are both drawn from $\mathcal{U}(0,2\pi)$. Lastly, the orbital eccentricities of the planets in each realization are treated as free parameters and are drawn from $\mathcal{U}(0,0.3)$ where the upper eccentricity limit was chosen as any orbit that is initialized with $e\gtrsim 0.3$ will undergo an immediate orbit crossing in less than one orbital timescale. 

Each simulated planetary system is integrated forward in time until one of the following stopping conditions is reached:

\begin{enumerate}
    \item Any pair of planets come within one mutual Hill radius:
    \begin{equation}
        R_{\text{Hill}} = \left( \frac{m_{p,i}+m_{p,i+1}}{3M_s} \right)^{1/3} \frac{a_i+a_{i+1}}{2}.
    \end{equation} 
    \item Any planet that travels beyond the imposed maximum barycentric distance of 0.2 AU ($\sim 4 a_{p,d}$). 
    \item The integration reaches its stopping time of $10^6$ orbits of the outermost planet; $\sim 2\times 10^4$ years. 
\end{enumerate}

\noindent Simulations that are halted because of either of the former two stopping criteria are flagged as unstable systems and the corresponding initial eccentricities are ruled out due to instability. The remaining systems that survive until the end of the simulation are deemed stable. Note that due to the short duration of the simulations performed here compared to the expected age of L 98-59 \citepalias[$>1$ Gyr;][]{kostov19}, these simulations are not intended to provide a detailed overview of the system's long-term stability but instead are used solely for the purpose of constraining the planetary eccentricities beyond that which can be measured by the RV data alone.

\begin{figure}
    \centering
    \includegraphics[width=\hsize]{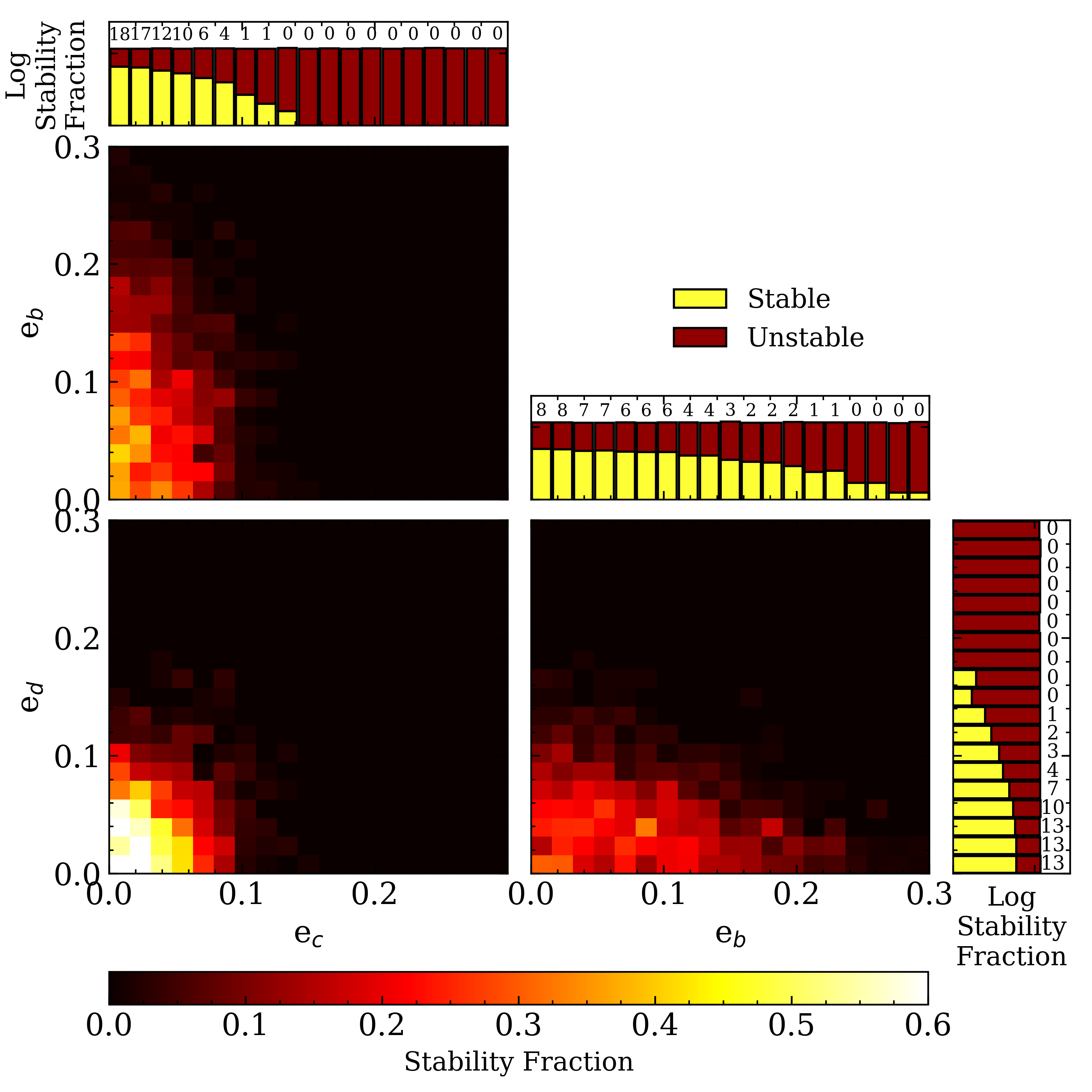}
    \caption{Stability maps of the L 98-59 three planet system as functions of the planet's initial orbital eccentricities. The 2D maps depict the fraction of stable systems computed from a set of N-body integrations using the planetary masses measured in this study. The 1D histograms depict the number of stable and unstable systems as a function of each planet's eccentricity separately and marginalized over all other dynamical parameters. The histograms are depicted on a logarithmic scale. The annotated numbers indicate each histogram bin's stability fraction in percentages.}
    \label{fig:stable}
\end{figure}

The fraction of stable systems as a function of each planet's initial orbital eccentricity is shown in Fig.~\ref{fig:stable}. The strong stability constraints on each planet's eccentricity are evident as the majority of the 3-parameter space exhibits a stability fraction that is consistent with zero. In particular,
for the two more massive planets in the system (i.e. L 98-59c and d), planetary systems for which $e_c$ or $e_d \gtrsim 0.1$ have a stability fraction of $<1$\%. Conversely, a small fraction of planetary systems with $0.1 \leq e_b \lesssim 0.2$ can remain stable as the median mass of L 98-59b is only $\sim 16$\% of either of the other two planets and therefore its eccentricity has a reduced effect on the overall stability of the system \citep{barnes06}. 

\begin{figure}
    \centering
    \includegraphics[width=.9\hsize]{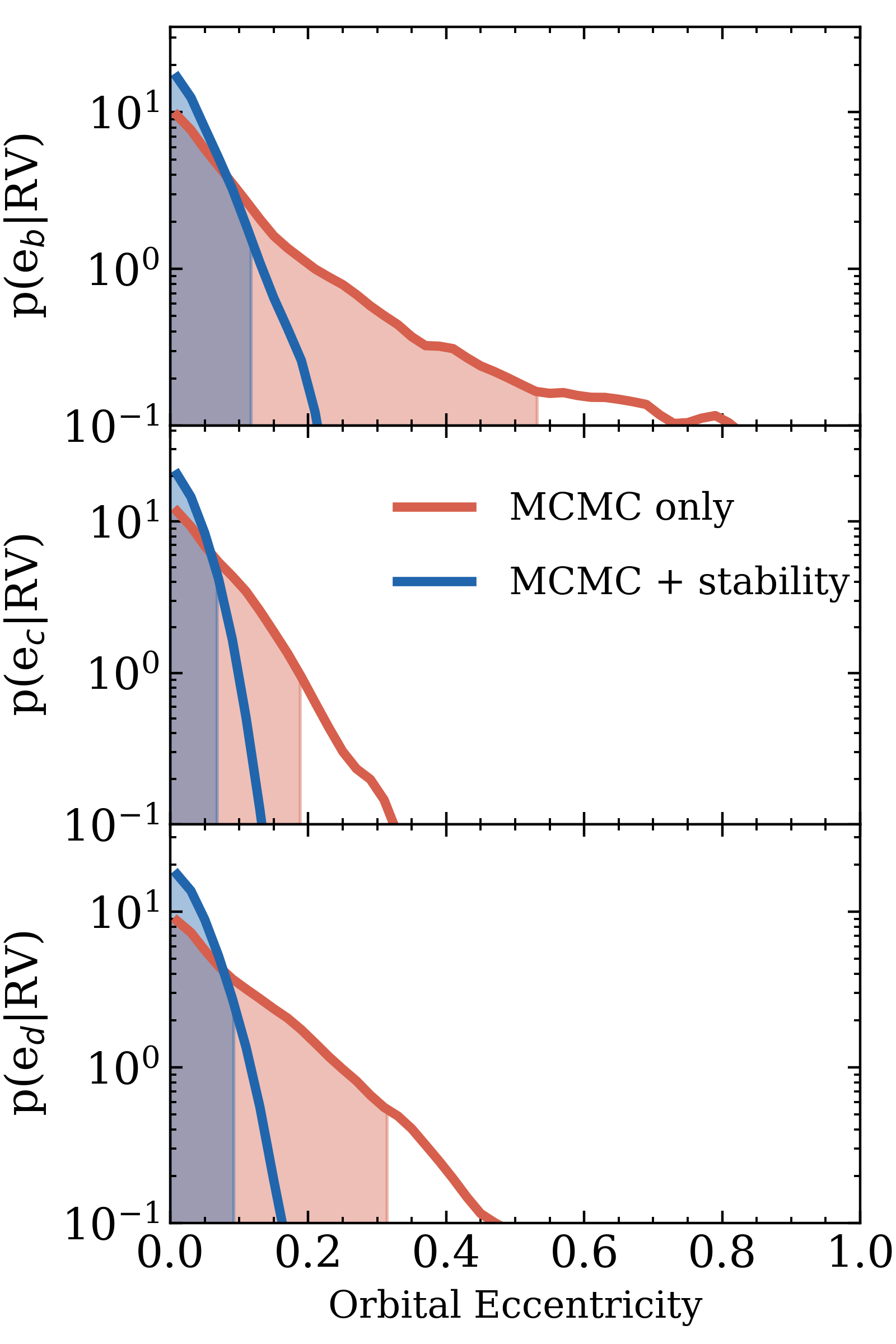}
    \caption{Marginalized posterior PDFs of the orbital eccentricities of each L 98-59 planet plotted on a logarithmic scale for improved visibility. The broader \emph{red} PDF for each planet corresponds to the MCMC only results representing the eccentricity constraint from the RV data alone. The shallower \emph{blue} PDFs combine the MCMC results with an additional prior from stability and therefore provides a stronger constraint on each planet's eccentricity. The shaded regions highlight the 95\% confidence intervals.}
    \label{fig:ecc}
\end{figure}

The large fraction of initial eccentricity values that result in an unstable orbital configuration as seen in Fig.~\ref{fig:stable}, provides constraints on the planets' orbital eccentricity values. If initially we ignore the fractional stability criteria derived from our dynamical simulations, we can derive posterior PDFs of the orbital eccentricities $e_i$ from the PDFs of $h_i$ and $k_i$ obtained from our MCMC analysis (Fig.~\ref{fig:corner}). The resulting $e_i$ posteriors are depicted in Fig.~\ref{fig:ecc} and represent our measurements of $e_i$ from the RV data alone. Next, we treat the stability fraction as a function of each $e_i$ as an additional prior on $e_i$ and resample the $e_i$ posterior PDFs according to the stability fraction. That is, for each sample from the joint $\{e_b,e_c,e_d \}$ posterior derived from MCMC, the probability that that sample is retained is given by the stability fraction of simulated planetary systems with those eccentricity values $\pm 0.02$. In this way high eccentricity values that cannot be ruled out by the MCMC analysis alone are frequently rejected because they often result in an unstable orbital configuration. 

The resampled $e_i$ posteriors that account for system stability are compared to the MCMC only results in Fig.~\ref{fig:ecc}. The distinct narrowing of each $e_i$ posterior after including the stability criteria indicates that the joint RV $+$ stability data provide the strongest constraints on the orbital eccentricities of the planets in the compact L 98-59 system. From each set of $e_i$ posteriors we derive eccentricity upper limits at 95\% confidence. The resulting upper limits from the RV data alone are $e_b < 0.53$, $e_c<0.19$, and $e_d < 0.31$. By comparison, the inclusion of the stability criteria results in drastically improved upper limits of $e_b < 0.12$, $e_c<0.07$, and $e_d < 0.09$. These measurements confirm that the planets in the L 98-59 compact planetary system all likely have eccentricities $\lesssim 0.1$, a result that is consistent with similarly compact systems exhibiting low eccentricities \citep{hadden14,vaneylen15} and the low dispersion in mutual inclinations in the system \citepalias[$\Delta i\sim 0.4^{\circ}$;][]{kostov19}.

\section{Discussion \& Conclusions}
\label{sec:conclusion}
In this study we conducted an intensive HARPS RV follow-up campaign of the L 98-59 multi-planet system to characterize the masses of its three known transiting planets \citepalias{kostov19}. We measure planet masses of the two outermost planets of $m_{p,c}=2.46\pm 0.31$ M$_{\oplus}$ and $m_{p,d}=2.26\pm 0.50$ M$_{\oplus}$ and derive an upper limit on the mass of the innermost planet of $m_{p,b}<0.98$ at 95\% confidence. The resulting bulk densities of the two outer planets are roughly consistent with bulk terrestrial compositions (see Fig.~\ref{fig:MR}) while the small size and small orbital separation place the innermost planet interior to the photoevaporation valley \citep{owen13,jin14,lopez14,chen16,lopez16,owen17} thus providing supporting evidence for its terrestrial nature as well. Confirmation of the terrestrial nature of L 98-59b will likely require $\mathcal{O}(500)$ additional RVs with a similar level of precision as our HARPS RVs to measure the semi-amplitude of L 98-59b at $3\sigma$ \citep{cloutier18b} given its expected semi-amplitude of 0.32 \mps{} \citep{chen17}. 

With the precise RV planet masses presented in this study, L 98-59c and d add to the growing list of planets to directly contribute to the completion of the TESS level one science requirement of delivering the masses of fifty planets smaller than 4 R$_{\oplus}$. Furthermore, at 1.35 and 1.57 R$_{\oplus}$ respectively, L 98-59c and d are among the smallest TESS planets to have precisely measured masses via RV  follow-up observations.

The nearby L 98-59 system of three terrestrial planets in a compact configuration within 7.5 days presents an ideal opportunity for comparative atmospheric planetology. To quantify the feasibility of detecting atmospheric signatures from the L 98-59 planets in transmission using JWST (assuming cloud-free atmospheres), we compute the \emph{transmission spectroscopy metric} from \cite{kempton18} using the MAP planet parameters measured in this study. We find that TSM$_b > 14.6$\footnote{Assuming that 95\% upper limit on $m_{p,b}$ from Table~\ref{tab:results}.}, TSM$_c=23.6\pm 5.2$, and TSM$_d=212\pm 76$. For L 98-59c and d this amounts to $\sim 0.8-1.3$ and $\sim 6-13$ times that of GJ 1132b \citep{dittmann17a,bonfils18}, the previously `best' prospect for the atmospheric characterization of a terrestrial-sized exoplanet from the pre-TESS era \citep{morley17}. Hence L 98-59c, and particularly L 98-59d, represent extremely promising targets for the atmospheric characterization of hot terrestrial exoplanets.

Similarly we compute the \emph{emission spectroscopy metric} from \cite{kempton18} assuming that the planet day-side temperatures are equivalent to their equilibrium temperatures assuming an Earth-like albedo (i.e. $A=0.3$). We find ESM$_b=2.2\pm 0.4$, ESM$_c=3.4\pm 0.6$, ESM$_d=1.5\pm 0.4$ such that the L 98-59 planets are somewhat less favorable for emission spectroscopy characterization compared to GJ 1132b (ESM$_{\text{GJ1132b}}=3.6\pm 0.5$) although L 98-59c still represents a viable target for such observations. The disfavourability of the L 98-59 planets compared to GJ 1132b is largely due to the larger stellar radius of L 98-59. Nevertheless, the close proximity of L 98-59 ($d=10.6$ pc) continues to make each of its known planets viable candidates for the characterization of hot terrestrial exoplanet atmospheres in emission if it can first be demonstrated on a slightly more favourable target such as GJ 1132b.

Regarding the prospect of the direct detection of the L 98-59 planets in reflected light using near-IR imagers on-board the next generation of extremely large telescopes, the planets' small angular separations ($\theta_b=0.0021'', \theta_c=0.0030'', \theta_d=0.0048''$) make them difficult to resolve despite having modest planet-star contrasts ($0.6,1.0,0.5 \times 10^{-6}$ respectively). Thus despite its close proximity, the orbital architecture of the known planets around L 98-59 is likely too compact for any of the planets to be directly imagable within the next decade.

Lastly, we emphasize that L 98-59 is slated to be re-observed by TESS within sectors 5, 8, 9, 10, 11, and 12. The extended baseline beyond the single 27 day field from sector 2 will provide opportunities to improve the orbital ephemerides and radii of the known planets and to continue to search for TTVs. If detected, TTV measurements could enable independent measurements of the planet masses for direct comparison to the RV results presented herein. The extended observational baseline may also enable the detection of additional planets at long orbital periods although the RVs presented herein do not show any significant evidence for such planets.

\begin{acknowledgements}
  R.C. is supported in part by the Natural Sciences and Engineering Research Council of Canada and 
  acknowledges that this work was performed on land traditionally inhabited by the Wendat, the Anishnaabeg, Haudenosaunee, Metis, and the
  Mississaugas of the New Credit First Nation.
  N.A-D. acknowledges the support of FONDECYT project 3180063. 
  X.B. acknowledges funding from the European Research Council under the ERC Grant Agreement n. 337591-ExTrA.
  J.S.J acknowledges support by FONDECYT grant 1161218 and partial support from CONICYT project Basal AFB-170002.
  This work is supported by the French National Research Agency in the  framework of the Investissements d’Avenir program (ANR-15-IDEX-02), through the funding of the ``Origin of Life'' project of the Univ. Grenoble-Alpes. 
  N.C.S was supported by FCT/MCTES through national funds and by FEDER - Fundo Europeu de Desenvolvimento Regional through COMPETE2020 - Programa Operacional Competitividade e Internacionalização by these grants: UID/FIS/04434/2019; PTDC/FIS-AST/32113/2017 \& POCI-01-0145-FEDER-032113; PTDC/FIS-AST/28953/2017 \& POCI-01-0145-FEDER-028953.
  Funding for the TESS mission is provided by NASA's Science Mission directorate.
  We acknowledge the use of public TESS Alert data from pipelines at the TESS Science Office and at the TESS Science Processing Operations Center.
  This research has made use of the Exoplanet Follow-up Observation Program website, which is operated by the California Institute of Technology, under contract with the National Aeronautics and Space Administration under the Exoplanet Exploration Program.
  Resources supporting this work were provided by the NASA High-End Computing (HEC) Program through the NASA Advanced Supercomputing (NAS) Division at Ames Research Center for the production of the SPOC data products.
  This paper includes data collected by the TESS mission, which are publicly available from the Mikulski Archive for Space Telescopes (MAST).
\end{acknowledgements}

\bibliographystyle{aa} % style aa.bst
\bibliography{refs} % your references Yourfile.bib

\end{document}